\documentclass[10pt,conference]{IEEEtran}

\IEEEoverridecommandlockouts
\usepackage{cite}
\usepackage{amsmath,amssymb,amsfonts}
\usepackage{algorithmic}
\usepackage{graphicx}
\usepackage{textcomp}
\usepackage{xcolor}
\usepackage{enumitem}
\usepackage{makecell}
\usepackage{threeparttable}
\usepackage{multirow}
\usepackage{fontawesome}
\usepackage{subfigure}
\usepackage{verbatim}
\usepackage{balance}
\usepackage{pifont}
\usepackage{adjustbox}
\usepackage{rotating}
\usepackage[hidelinks]{hyperref}
\usepackage{booktabs}
\usepackage{soul}
\usepackage{fancyhdr} 
\usepackage{inconsolata}
\usepackage{tcolorbox}
\usepackage[lined,boxed,linesnumbered,commentsnumbered,ruled]{algorithm2e}
\usepackage{color}
\usepackage{arydshln}
\usepackage{colortbl}
\definecolor{mygray}{gray}{.9}

\newcommand{\tech}{$\mu$FiX}
\newcommand{\techVariantWithoutSelfimprovement}{\tech{}$_{\textit{woS}}$}
\newcommand{\techVariantWithoutFeedback}{\tech{}$_{\textit{woF}}$}
\newcommand{\techVariantSCoT}{\tech{}$_{\textit{SCoT}}$}
\newcommand{\techVariantSelfrepair}{\tech{}$_{\textit{SR}}$}
\newcommand{\techVariantWithoutTestcase}{\tech{}$_{\textit{woT}}$}

\newcommand{\xz}[1]{{\color{red}[XZ: #1]}}

\newcommand{\Comment}[1]{}

\newcommand{\revision}[1]{\textcolor{black}{#1}}
\newcommand{\checkedrevision}[1]{\textcolor{black}{#1}}

\def\BibTeX{{\rm B\kern-.05em{\sc i\kern-.025em b}\kern-.08em
    T\kern-.1667em\lower.7ex\hbox{E}\kern-.125emX}}

\newcommand{\distance}{4pt}
\begin{document}

\title{Fixing Large Language Models' Specification Misunderstanding for Better Code Generation}

\author{\IEEEauthorblockN{Zhao Tian}
\IEEEauthorblockA{\textit{College of Intelligence and} \\
\textit{Computing, Tianjin University}\\
China \\
tianzhao@tju.edu.cn}
\and
\IEEEauthorblockN{Junjie Chen*}
\IEEEauthorblockA{\textit{College of Intelligence and} \\
\textit{Computing, Tianjin University}\\
China \\
junjiechen@tju.edu.cn}
\and
\IEEEauthorblockN{Xiangyu Zhang}
\IEEEauthorblockA{\textit{Department of Computer Science,} \\
\textit{Purdue University}\\
USA \\
xyzhang@cs.purdue.edu}
\thanks{*Junjie Chen is the corresponding author.}
}

\maketitle

\thispagestyle{fancy}
\renewcommand{\headrulewidth}{0pt}
\renewcommand{\footrulewidth}{0pt} 
\pagestyle{fancy}

\begin{abstract}
Code generation is to automatically generate source code conforming to a given programming specification, which has received extensive attention especially with the development of large language models (LLMs).
Due to the inherent difficulty of code generation, the code generated by LLMs may not be aligned with the specification.
Although thought-eliciting prompting techniques have been proposed to enhance the code generation performance of LLMs, producing correct understanding for complicated programming problems remains challenging, resulting in unsatisfactory performance.
Also, some feedback-based prompting techniques have been proposed to fix incorrect code using error messages produced by test execution.
However, when the generated code deviates significantly from the ground truth, they encounter difficulties in improving performance based on such coarse-grained information.

In this work, we propose a novel prompting technique, called \tech{}, to improve the code generation performance of LLMs by devising both sophisticated thought-eliciting prompting and feedback-based prompting and making the first exploration on their synergy.
It first exploits test case analysis to obtain specification understanding and enables a self-improvement process to identify and refine the misunderstanding in the thought-eliciting prompting phase.
\tech{} further fixes the specification understanding towards the direction reducing the gap between the provided understanding (from the first phase) and the actual understanding implicitly utilized by LLMs for code generation in the feedback-based prompting phase.
By \checkedrevision{improving the understanding} with \tech{}, the code generation performance of LLMs can be largely improved.
Our evaluation on two advanced LLMs (ChatGPT and DeepSeek-Coder) with six widely-used benchmarks by comparing with 15 baselines, demonstrates the effectiveness of \tech{}.
For example, \tech{} outperforms the most effective baseline with an average improvement of 35.62\% in terms of Pass@1 across all subjects.
\end{abstract}


\vspace{-1mm}
\vspace{-1mm}
\section{Introduction}
\label{sec:introduction}
Code generation aims to automatically generate source code conforming to a given programming specification (which is usually a natural language description).
It can help reduce repetitive programming efforts and improve software development productivity.
In recent years, code generation has received extensive attention from both academia and industry.
In particular, with LLMs being rapidly developed, significant progress has been made in code generation, such as ChatGPT~\cite{chatgpt2022} and DeepSeek-Coder~\cite{deepseekcoder2023}.
The LLMs take the programming specification (i.e., prompt) as input and output the corresponding code solution, demonstrating notable advancements in code generation.

Despite their popularity, LLMs still suffer from some performance issues.
That is, the generated code may be not aligned with the human-provided specification, especially when the programming logic is complicated~\cite{ma2023bridging}.
For example, one state-of-the-art LLM, i.e., ChatGPT, generates code passing all test cases for only 16.33\% of programming problems in a real-world benchmark~\cite{hendrycks2021measuring}.
The performance issues can negatively affect the practical use of LLMs, even slow down software development process and harm software quality.
Hence, it is important to enhance the ability of LLMs in code generation.

Although fine-tuning strategies have been widely adopted to improve the code generation performance of LLMs, they are time-consuming and require a large amount of computing resources~\cite{kaddour2023challenges,tian2023code,tian2023fly,tian2024large}.
In recent years, prompting techniques have been proposed to achieve this goal in a plug-and-play manner~\cite{wei2022chain,dong2023self,liu2023improving,nashid2023retrieval,jiang2023self}.
Among them, thought-eliciting prompting is the most popular category.
It aims to elicit LLMs to produce intermediate reasoning steps as the specification understanding for more accurate code generation.
The typical thought-eliciting prompting techniques include CoT~\cite{wei2022chain} (that elicits LLMs to produce intermediate natural language reasoning steps),
Self-planning~\cite{jiang2023self} (that guides LLMs to decompose the specification into a set of easy-to-solve sub-problems and produce code plans to facilitate code generation),
SCoT~\cite{li2023enabling} (that enhances CoT by utilizing program structures to build intermediate reasoning steps, thereby eliciting LLMs for code generation), and so on.
Feedback-based prompting is another category of prompting techniques (such as Self-Debugging~\cite{chen2023teaching}, Self-Edit~\cite{zhang2023self}, and Self-repair~\cite{olausson2023demystifying}), which leverages error messages produced by test execution to enable LLMs to fix incorrectly generated code.

Although these techniques have been studied extensively, their performance still needs to be improved.
For example, the existing thought-eliciting prompting techniques have difficulties in producing correct understanding according to the (concise) specification, when facing with complicated programming problems~\cite{jiang2023self,mu2023clarifygpt}.
Moreover, they cannot identify or fix incorrect specification understanding that occurs before code generation, leading to inaccurate results.
For feedback-based prompting, the existing techniques just utilize error messages produced by test execution to understand why the generated code is incorrect, which is too coarse-grained to identify root causes and thus leads to suboptimal performance.
If the generated code deviates largely from the ground truth, it is quite hard for them to improve the performance based on error messages from such low-quality generated code~\cite{olausson2023demystifying}.
In particular, the two categories play roles at different stages (the former working before test execution whereas the latter after)
but there is no existing work exploring their synergy.

To improve the code generation performance of LLMs, we propose a novel prompting technique, called \textbf{\tech{}} 
(\textbf{M}is\textbf{u}nderstanding \textbf{FiX}ing), 
to overcome the aforementioned limitations. 
In particular, \tech{} is the first to explore the synergy of the above two categories, by devising both thought-eliciting prompting and feedback-based prompting.
The key insight is that, by \checkedrevision{improving specification understanding} in the thought-eliciting prompting phase via test case analysis, the effectiveness of the subsequent feedback-based prompting (and code generation) can be enhanced.
Namely, even though LLMs may not generate correct code based on the specification understanding, it can make the generated code as close to the ground truth as possible, which can provide a greater chance for the subsequent feedback-based prompting to further improve the code generation performance.

In the thought-eliciting prompting phase, the key lies in \checkedrevision{improving LLMs' understanding} of the specification. 
We propose to achieve this by leveraging test cases inherently included as part of a specification. 
Note that such test cases are prevalent in practice, including those widely-studied datasets in code generation (e.g., HumanEval~\cite{chen2021evaluating} and APPS~\cite{hendrycks2021measuring}). 
Due to the intrinsic reasoning capabilities of LLM, if its understanding is correct, it shall be able to resolve a natural language request derived from the (embedded) test cases and produce expected outputs. 
\checkedrevision{In contrast, when the understanding is incorrect, by prompting LLM to fix the test outputs, we could improve the understanding}, thereby enhancing the later code generation performance. 
This is analogous to how software developers use test case examples to understand complicated programming logic in practice~\cite{beck2003test,gulwani2016programming}.

Although the understanding can be refined to infer the correct test outputs, it does not mean that the corresponding generated code can really pass the test cases in actual execution due to the gap between specification understanding and code generation.
Specifically, specification understanding and code generation emphasize different aspects of abilities in LLMs, where the former relies on the reasoning ability of LLMs while the latter emphasizes the ability of translating the natural language description into the source code~\cite{roziere2023code,guo2024deepseek}.
Hence, when invoking LLMs to generate code, the actual understanding implicitly utilized by LLMs may be inconsistent with the provided understanding.
\tech{} further designs feedback-based prompting to improve code generation performance (if the test execution fails).
Instead of prompting LLMs to directly fix the generated code with coarse-grained error messages,
\tech{} prompts LLMs to understand the root cause (i.e., the above-mentioned gap) by comparatively analyzing the provided understanding and the actual understanding (obtained via code summarization~\cite{ahmed2024automatic}), and then adjust the natural language description for the specification understanding to enhance its accessibility to the code generation ability of LLMs.
It is also a kind of fixing on specification understanding, which aims to make the understanding be better utilized by the code generation ability of LLMs.

We conducted extensive experiments to evaluate \tech{} on two state-of-the-art LLMs (i.e., ChatGPT~\cite{chatgpt2022} and DeepSeek-Coder~\cite{deepseekcoder2023}) based on six widely-used benchmarks.
Our results show that \tech{} significantly outperforms all the 15 compared prompting techniques on both LLMs across all six benchmarks, demonstrating our idea for enhancing code generation performance by \checkedrevision{improving LLMs' specification understanding}.
For example, the average improvement of \tech{} over all 15 compared techniques is 35.62\%$\sim$80.11\% in terms of Pass@1 (measuring the ratio of programming problems for which the generated code passes all evaluation test cases) across all subjects.
Moreover, we constructed four variants of \tech{} for the ablation study.
The results confirm the contribution of our thought-eliciting prompting and feedback-based prompting strategies in \tech{}.

We summarize our contributions in this work as follows:
\begin{itemize}[leftmargin=10pt]

    \item We propose a novel prompting technique, called \tech{}, \checkedrevision{to enhance code generation performance of LLMs by improving their specification understanding} with both sophisticated thought-eliciting prompting and feedback-based prompting.

    \item We design a novel thought-eliciting prompting strategy, which exploits test case analysis to produce more accurate specification understanding and enables a self-improvement process to identify and \checkedrevision{refine} the misunderstanding before generating code.

    \item We design a novel feedback-based prompting strategy that adjusts the understanding towards the direction reducing the gap between the provided understanding and the actual understanding implicitly utilized by LLMs for code generation.

    \item We conducted extensive experiments on two LLMs (i.e., ChatGPT and DeepSeek-Coder) across six widely-used benchmarks by comparing with 15 baselines, demonstrating the effectiveness of \tech{} for improving the code generation performance of LLMs.  
\end{itemize}

\section{An Illustrating Example}
\label{sec:example}

\begin{figure*}[t!]
    \centering
    \includegraphics[width=1.0\linewidth]{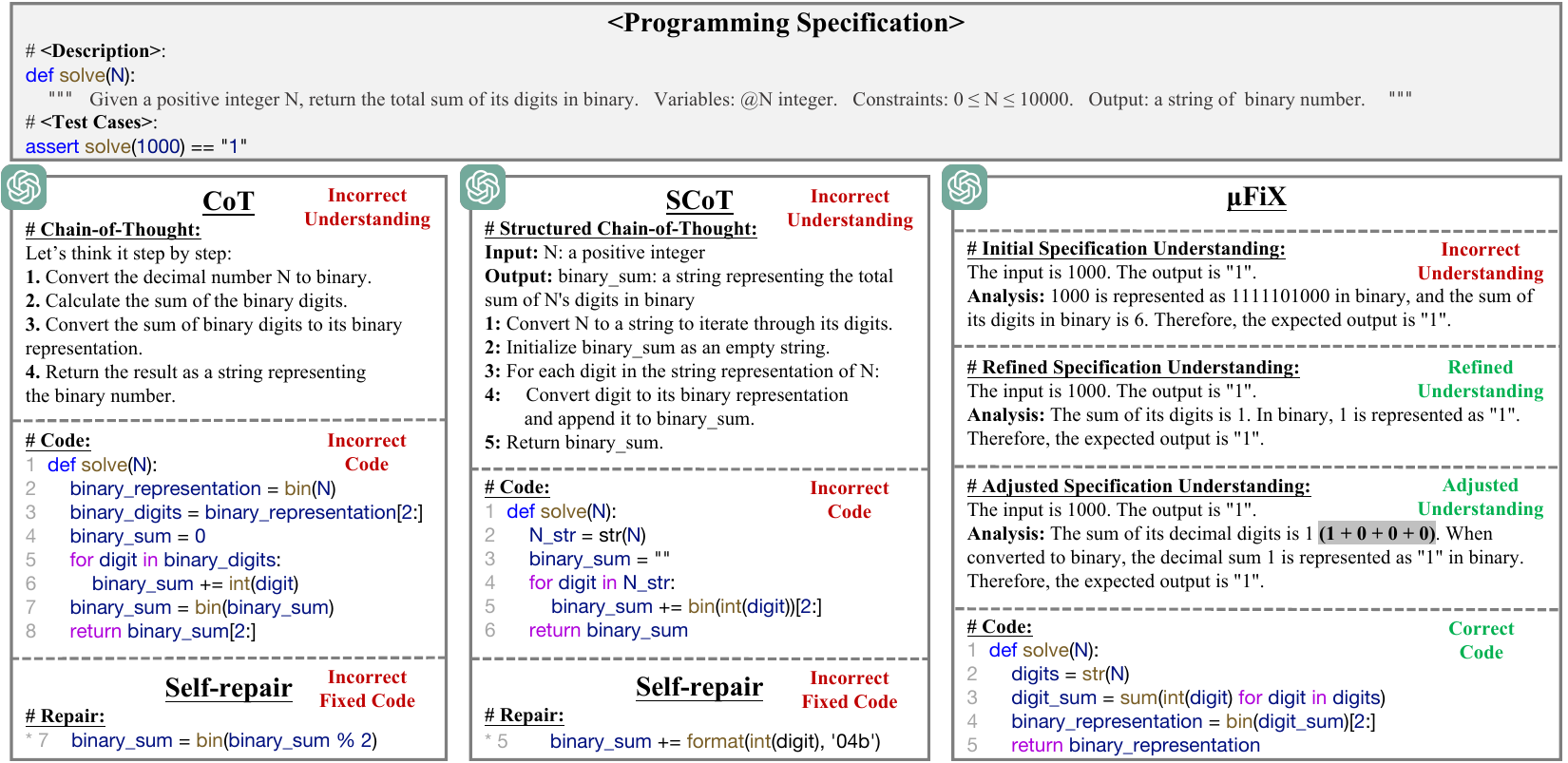}
    \vspace{-6mm}
    \caption{An example from HumanEval \#84 with ChatGPT}
    \label{fig:example}
\end{figure*}

We use an example to illustrate why the state-of-the-art prompting techniques do not work well, motivating our idea.
Figure~\ref{fig:example} shows the (simplified) example sampled from the real-world HumanEval benchmark~\cite{chen2021evaluating}.
In this figure, the programming specification provides a natural language description (including a test case).

Recent work has shown that prompting techniques can obtain the understanding of the specification by exploiting the logical reasoning ability of LLMs, thereby improving their performance in code generation~\cite{jiang2023self,mu2023clarifygpt}.
Here, we employed two state-of-the-art thought-eliciting prompting techniques (i.e., CoT~\cite{wei2022chain} and SCoT~\cite{li2023enabling}) to prompt the state-of-the-art LLM (i.e., ChatGPT~\cite{chatgpt2022}) for generating code with regard to this specification.
As shown in Figure~\ref{fig:example}, both CoT and SCoT obtained the understanding of the specification by dividing it into several intermediate reasoning steps (i.e., chain of thought and structured chain of thought).
Then, they integrated these detailed reasoning steps with the original specification as the prompt, facilitating ChatGPT for code generation.

However, we found that both obtained understandings are incorrect due to the complicated programming logic implied in the specification, leading to generating incorrect code.
Moreover, both CoT and SCoT are incapable of checking the correctness of the specification understanding before code generation, further limiting their performance.
\ul{This motivates the importance of 
improving specification understanding
before code generation for a thought-eliciting prompting technique.}

Subsequently, we employed the state-of-the-art feedback-based prompting technique, Self-repair~\cite{olausson2023demystifying}, to fix the incorrect code generated by CoT and SCoT.
It leverages error messages produced by test execution on the incorrect code to understand why it is incorrect and then prompts ChatGPT to fix the code accordingly.
Unfortunately, both fixed code remained incorrect, due to the substantial deviation of the initially generated code from the ground truth.
It is quite hard to improve the code generation performance solely based on error messages from such low-quality code. 
\ul{This motivates that a feedback-based prompting technique requires not only possessing the ability to improve incorrectly generated code but also starting with high-quality code.}

The two categories of techniques operate at different stages (the former working before test execution whereas the latter after). 
The output of thought-eliciting prompting (i.e., code generated by LLMs enhanced by the specification understanding) serves as the input of feedback-based prompting.
That is, the effectiveness of the former can affect that of the latter, thereby affecting the code generation performance, but there is no existing work exploring this synergy.
\ul{This motivates us to explore the synergy of these two categories for better code generation.}
Specifically, we can first \checkedrevision{improve the specification understanding} for generating the code as close to the ground truth as possible in the thought-eliciting prompting phase, and then design a more effective feedback-based prompting strategy for generating more accurate code on the basis of high-quality code from the first phase.
With this insight, we propose a novel prompting technique, called \tech{}, which leverages the test cases inherent in the specification to achieve this goal.
With \tech{}, the \checkedrevision{improved specification understanding} is indeed produced, and also the correct code is generated for this example in Figure~\ref{fig:example}.
\vspace{-1mm}
\section{Approach}
\label{sec:approach}
\vspace{-4mm}
\subsection{Overview}
\label{sec:overview}
We propose a novel prompting technique, called \textbf{\tech{}}, to improve the code generation performance of LLMs.
\tech{} devises both sophisticated thought-eliciting prompting and feedback-based prompting, as the first exploration on the synergy of both categories of prompting techniques.
The key insight is to utilize test cases (inherently included in specifications) to \checkedrevision{improve specification understanding}, inspired by the practice where software developers often use test cases to understand complex programming logic.
It is helpful to generate high-quality code (close to the ground truth) as the starting point for the feedback-based prompting phase to further improve LLM performance.
In practice, test cases are commonly included in programming specifications~\cite{chen2021evaluating,hendrycks2021measuring}, and we also evaluated the influence of the number of test cases on the effectiveness of \tech{} in Section~\ref{sec:influence_testcase}.
If a specification lacks test cases, existing work has demonstrated the effectiveness of LLMs in generating them~\cite{chen2022codet,liu2023your}, thereby ensuring the feasibility of \tech{}, as detailed in Section~\ref{sec:influence_testcase}.

\begin{figure*}[t!]
    \centering
    \includegraphics[width=1.0\linewidth]{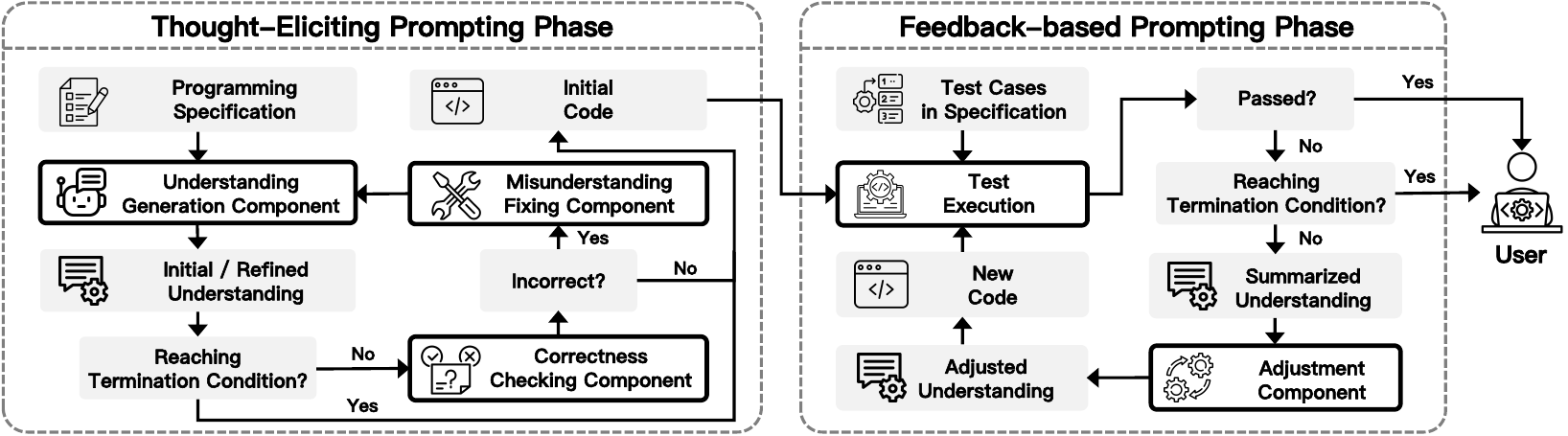}
    \vspace{-6mm}
    \caption{Overview of \tech{}}
    \label{fig:overview}
    \vspace{-1mm}
\end{figure*}

Figure~\ref{fig:overview} shows the overview of \tech{}. 
It consists of two phases: 
(1) the thought-eliciting prompting phase (Section~\ref{sec:phase_1}), which emphasizes the analysis on test cases for producing specification understanding and enables a self-improvement process to refine the misunderstanding with the aid of test cases,
and (2) the feedback-based prompting phase (Section~\ref{sec:phase_2}), which comparatively analyzes the provided understanding and the actual understanding implicitly utilized by LLMs for code generation and minimizes the gap between the two for better code generation.
The latter is activated only if the generated code does not pass the test cases (used for obtaining the specification understanding) in actual execution.

For ease of understanding on our \tech{}, we reuse the example (i.e., HumanEval \#84) introduced in Section~\ref{sec:example} to illustrate the details of \tech{}.

\begin{figure}[t!]
    \centering
    \includegraphics[width=1.0\linewidth]{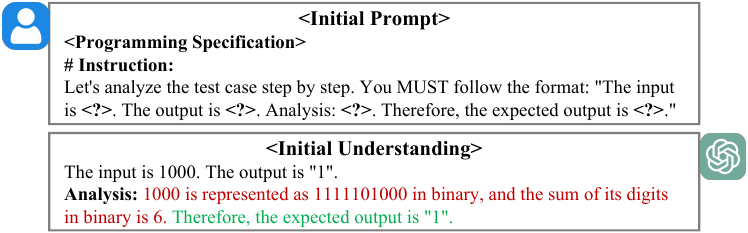}
    \vspace{-6mm}
    \caption{An example of \tech{} in understanding generation}
    \label{fig:example_tcot_1}
\end{figure}

\vspace{-1mm}
\subsection{Thought-Eliciting Prompting Phase}
\label{sec:phase_1}
In the thought-eliciting prompting phase, \tech{} takes the programming specification as input, and aims to output \checkedrevision{more accurate specification understanding by leveraging the test cases inherently included in the specification}.
On one hand, emphasizing test case analysis helps LLMs \checkedrevision{improve} specification understanding because test cases contain more specific details that facilitate the understanding of complicated programming logic.
On the other hand, \checkedrevision{test cases enable the self-improvement process to refine LLMs' misunderstanding} before generating code.
In the self-improvement process, \tech{} first checks the correctness of the specification understanding and then \checkedrevision{improves} it (if it is a misunderstanding).
Note that existing thought-eliciting prompting techniques cannot identify and refine misunderstanding before generating code due to overlooking the importance of such test cases.
Overall, the thought-eliciting prompting phase in \tech{} consists of three steps: initial understanding generation, correctness checking of the understanding, and misunderstanding fixing.
These steps will be detailed in the following.

\smallskip
\underline{\textit{Initial understanding generation}}:
As shown in Figure~\ref{fig:example}, existing thought-eliciting prompting techniques also take the programming specification (including test cases) as input, but they do not pay special attention to these test cases, thereby limiting their effectiveness.
Instead, \tech{} emphasizes test case analysis by providing a structured instruction to LLMs, which can help them utilize specific test cases to understand complicated programming logic.
The {\tt <Initial Prompt>} in Figure~\ref{fig:example_tcot_1} shows the structured instruction activating test case analysis, which facilitates explaining the programming logic that processes the inputs provided in the test cases to get the expected output specified in the test cases.
Then, \tech{} enables LLMs to complete {\tt <?>} in the instruction, producing initial specification understanding.
For example, Figure~\ref{fig:example_tcot_1} ({\tt <Initial Understanding>}) shows the initial understanding produced by ChatGPT with test case analysis regarding the programming specification in Figure~\ref{fig:example}.

\begin{figure}[t!]
    \centering
    \includegraphics[width=1.0\linewidth]{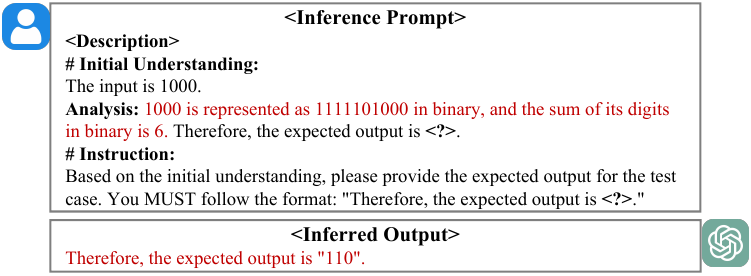}
    \vspace{-6mm}
    \caption{An example of \tech{} in correctness checking of the understanding}
    \label{fig:example_tcot_2}
\end{figure}

\smallskip
\underline{\textit{Correctness checking of the understanding}}: 
From Figure~\ref{fig:example_tcot_1}, we find that the initial understanding is incorrect even though the expected output shown in the understanding is correct.
Actually, the expected output is correctly specified due to its leakage in the test cases as shown in the programming specification of Figure~\ref{fig:example}.
With this specification misunderstanding, LLMs hardly generate correct code.
Hence, with the aid of test cases, \tech{} includes a mechanism to check the correctness of the understanding in advance, in order to have a chance for \checkedrevision{improving the understanding} before generating code.
Specifically, \tech{} masks the expected output in the understanding and enables LLMs to infer it according to the analyzed logic in the understanding.
If the inferred output differs from the expected output, it indicates an incorrect specification understanding.
As shown in Figure~\ref{fig:example_tcot_2}, this mechanism is effective to identify the misunderstanding (produced in Figure~\ref{fig:example_tcot_1}).

\smallskip
\underline{\textit{Misunderstanding fixing}}:
Once \tech{} identifies a specification misunderstanding, it prompts LLMs to refine this misunderstanding by providing the corresponding instruction (e.g., ``The above understanding is incorrect. Please provide the correct analysis...''). 
This process terminates until the refined understanding passes the checking mechanism or the number of refinements reaches the pre-defined threshold (denoted as $N$).
Note that passing the checking mechanism cannot guarantee that LLMs completely understand the programming specification as the number of test cases equipped in the specification tend to be limited.
However, at least, the refined understanding (as shown in {\tt Refined Specification Understanding} of Figure~\ref{fig:example}) passing the checking mechanism is more accurate than that does not pass it.
With such high-quality refined understanding, LLMs may directly generate correct code.
If not, \tech{} still provides a high-quality starting point for its subsequent feedback-based prompting phase for further improving code generation performance.

\vspace{-1mm}
\subsection{Feedback-based Prompting Phase}
\label{sec:phase_2}
By integrating the more accurate specification understanding into the programming specification along with an additional instruction ``Please implement the function in a markdown-style code block (pay attention to the specification understanding)'', \tech{} prompts LLMs to generate code.

Although the understanding produced in the thought-eliciting prompting phase is \checkedrevision{more accurate} regarding the test cases, it does not guarantee that the corresponding generated code will pass the test cases in actual execution.
This is due to the gap between specification understanding and code generation, which emphasize different aspects of abilities in LLMs.
That is, the actual understanding implicitly utilized by LLMs for code generation may be inconsistent with the provided understanding.
For example, when prompting LLMs to generate code, they may miss some important contents in the provided understanding due to the natural-language-description style unfamiliar to this ability of LLMs.

\begin{figure}[t!]
    \centering
    \includegraphics[width=1.0\linewidth]{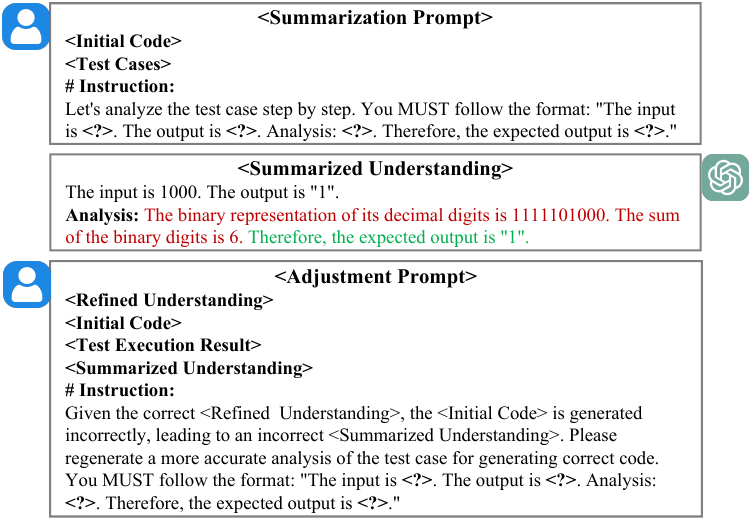}
    \vspace{-6mm}
    \caption{An example of \tech{} in adjusting the understanding. For saving space, we show the refined understanding and adjusted understanding in {\tt Refined Specification Understanding} and {\tt Adjusted Specification Understanding} of Figure~\ref{fig:example}, respectively.}
    \label{fig:example_tcot_4}
\end{figure}

Existing feedback-based prompting techniques leverage error messages produced by test execution to prompt LLMs for enhancing code generation~\cite{olausson2023demystifying}.
This kind of information is too coarse-grained to identify the root cause of incorrect code (especially when the generated code significantly deviates from the ground truth), leading to unsatisfactory performance.
Different from them, \tech{} prompts LLMs to understand the root cause (i.e., the aforementioned gap) by comparatively analyzing the provided \checkedrevision{refined understanding} and the actual understanding implicitly utilized by LLMs for code generation.
Here, based on the symmetry between code generation (from natural language descriptions to code) and code summarization~\cite{ahmed2022few,lu2021codexglue} (from code to natural language descriptions), \tech{} estimates the actual understanding implicitly utilized by LLMs for code generation through code summarization (also called summarized understanding for ease of presentation).
As shown in {\tt <Summarization Prompt>} of Figure~\ref{fig:example_tcot_4}, \tech{} prompts LLMs to produce the summarized understanding ({\tt <Summarized Understanding>} of Figure~\ref{fig:example_tcot_4}).

\tech{} then adjusts the natural language description of the \checkedrevision{refined specification understanding} towards the direction reducing the gap, by exploiting the logical reasoning ability of LLMs.
For example, some important contents missed by the summarized understanding can be highlighted or the unfamiliar natural-language-description style to LLMs can be improved accordingly through comparative analysis.
Here, we design an adjustment prompt ({\tt <Adjustment Prompt>} of Figure~\ref{fig:example_tcot_4}) to enable LLMs to achieve the above goal, which consists of the generated code, test execution results, the refined understanding produced from the thought-eliciting prompting phase, summarized understanding, and an additional instruction guiding LLMs for adjusting understanding. 

Taking {\tt Refined Specification Understanding} of Figure~\ref{fig:example} (\checkedrevision{the provided refined understanding}) and {\tt <Summarized Understanding>} of Figure~\ref{fig:example_tcot_4} (the summarized understanding) as an example for illustration, the comparative analysis between them implies that LLMs just capture some keywords (e.g., \textit{digits}, \textit{binary}, and \textit{sum}) but fail to understand the logic among keywords in the summarized understanding, which provides a direction to adjust the understanding.
Indeed, the adjusted understanding emphasizes the correct computation logic with {\tt \textbf{(1+0+0+0)}}, as shown in {\tt Adjusted Specification Understanding} of Figure~\ref{fig:example}.
With the adjusted understanding by \tech{}, ChatGPT successfully generates correct code for the programming specification, as shown in Figure~\ref{fig:example}.
This phase is an iterative process and terminates until the generated code passes the corresponding test cases or the number of adjustments reaches the threshold (denoted as $M$).

In fact, the adjustment process can be also regarded as a kind of fixing on specification understanding, which aims to ensure that the understanding is being correctly utilized during the code generation.
That is, the two phases in \tech{} involve complementary aspects of specification understanding fixing, and synergetically enhance code generation performance. 
\checkedrevision{
Note that both fixing components are not simple regeneration. 
Similar to human-centric iterative improvement processes (e.g., debugging), they essentially provide counter-factual evidence (e.g., historical incorrect cases and their explanations) as feedback to LLMs. 
Analogous to human processes, they may not succeed potentially due to the limited ability of LLMs or insufficient feedback to LLMs.
In the future, we will explore more effective LLMs in this aspect and more informative feedback so as to further enhance both fixing components. 
}

\section{Evaluation Design}
\label{sec:evaluation_design}
\vspace{-1mm}
Our study addresses the following research questions (RQs). 
\begin{itemize}
    \item \textbf{RQ1}: How does \tech{} perform in improving the code generation performance of LLMs compared to the state-of-the-art prompting techniques?

    \item \textbf{RQ2}: Does each main component in \tech{} contribute to the overall effectiveness?

    \item \checkedrevision{\textbf{RQ3}: How does \tech{} perform under different hyper-parameters' configurations?}
\end{itemize}

\vspace{-2mm}
\subsection{Studied LLMs}
\label{sec:llms}
\vspace{-1mm}
\checkedrevision{
Following the OpenAI's news~\cite{openaikills2023}, Codex's access was discontinued and GPT-3.5 has been recommended instead, and thus we selected GPT-3.5 (ChatGPT~\cite{chatgpt2022}) as the representative commercial LLM in our study following the existing studies~\cite{jiang2023self,olausson2023demystifying,tian2023chatgpt}.
Here, we did not choose GPT-4 due to its high cost.
Following the existing studies~\cite{zheng2024opencodeinterpreter,liu2024exploring,dou2024stepcoder}, we selected DeepSeek-Coder~\cite{guo2024deepseek} as the representative open-source LLM, since it has exhibited state-of-the-art effectiveness among open-source LLMs across multiple programming languages and various benchmarks in terms of coding capabilities~\cite{evalplusleaderboard2024,tabbymlteamleaderboard2024,testevalleaderboard2024}.}
Based on the two state-of-the-art LLMs in code generation, the generality of \tech{} can be investigated to some extent.
Specifically, we used ChatGPT (version gpt-3.5-turbo-0613) via OpenAI's APIs, and DeepSeek-Coder (version DeepSeek-Coder-6.7B-Instruct) from Huggingface~\cite{wolf2019huggingface}.
\checkedrevision{Note that LLM-generated code often includes natural-language text snippets, leading to compilation failures. Therefore, following the existing work~\cite{liu2023your}, we employed a code sanitizer tool~\cite{codesanitizer2023} to clean LLM-generated code.}

\vspace{-1mm}
\subsection{Benchmarks}
\label{sec:benchmarks}
To sufficiently evaluate \tech{}, we adopted six widely-used datasets in our study, i.e., HumanEval~\cite{chen2021evaluating}, HumanEval${+}$~\cite{liu2023your}, APPS~\cite{hendrycks2021measuring}, HumanEval-ET~\cite{dong2023codescore}, APPS-ET~\cite{dong2023codescore}, and MBPP-ET~\cite{dong2023codescore}.
These datasets have been widely used in many existing studies to measure the performance of prompting techniques in code generation~\cite{li2023enabling,li2023think,jiang2023self}.

\textit{\underline{HumanEval}} has 164 human-written programming problems proposed by OpenAI. 
The specification for each problem consists of a function signature, a natural language problem description, and several test cases.
Each problem has another set of test cases for evaluating the correctness of generated code, which is called \textit{evaluation test cases} here to distinguish with the test cases in the specification.

\textit{\underline{HumanEval${+}$}} extends HumanEval by automatically generating more evaluation test cases for each problem via the EvalPlus framework~\cite{liu2023your}.
More evaluation test cases could determine the correctness of generated code more accurately.

\textit{\underline{APPS}} contains 10,000 programming problems collected from public programming competition platforms (e.g., LeetCode~\cite{leetcode2023}), including 5,000 training data and 5,000 test data.
The specification for each problem contains a natural language problem description and several test cases.
Following the existing work~\cite{olausson2023demystifying}, we randomly sampled 300 problems from test data according to the difficulty distribution, so as to balance evaluation cost and conclusion generality.

The existing work~\cite{dong2023codescore} extended the HumanEval and APPS benchmarks as \textit{\underline{HumanEval-ET}} and \textit{\underline{APPS-ET}} by constructing over 100 additional evaluation test cases per problem, respectively.
These additional evaluation test cases cover more corner cases to enhance evaluation sufficiency on generated code.
Regarding APPS-ET, we used the same 300 problems as APPS.
Besides, we used \textit{\underline{MBPP-ET}}, an extended version of MBPP (that does not have additional evaluation test cases), by complementing evaluation test cases for each programming problem in a similar way.
This dataset contains 974 programming problems, each with a specification comprising a natural language problem description and three test cases.


Note that \tech{} just utilizes the test cases in the specification for each problem. 
For these datasets, 81.57\% of problems have at least three test cases in their corresponding specifications.
The evaluation test cases are just used for assessing generated code following the practice in code generation~\cite{dong2023codescore}.

\vspace{-2mm}
\subsection{Metrics}
\label{sec:metrics}
Following existing work~\cite{jiang2023self,dong2023codescore}, we executed evaluation test cases to check correctness of generated code for each programming problem, and calculated Pass@$k$ and AvgPassRatio to measure the performance of LLMs in code generation. 

\textit{\underline{Pass@$k$}} measures the functional correctness of generated code on evaluation test cases.
Given a programming problem, the LLM generates $k$ code instances. 
The problem is considered solved if any instance passes all evaluation test cases.
Pass@$k$ is the percentage of solved problems out of the total number of problems. 
As demonstrated by the existing work~\cite{dong2023self}, developers tend to consider and evaluate one code instance produced by the used LLM, and thus we set $k=1$ following the existing studies~\cite{yang2023chain,dong2023codescore,dong2023self,jiang2023self,mu2023clarifygpt}.
Note that Pass@1 is a more strict setting and thus improving it is challenging.
Larger Pass@$k$ values mean better code generation performance.

\textit{\underline{AvgPassRatio}} measures the degree of correctness of generated code on evaluation test cases, differing from Pass@$k$ that considers whether the generated code is completely correct on evaluation test cases~\cite{hendrycks2021measuring}.
Both metrics are complementary to a large extent.
AvgPassRatio calculates the ratio of passing evaluation test cases to all evaluation test cases for each problem, and then measures the average ratio across all problems.
Larger AvgPassRatio values mean better code generation performance.
For ease of presentation in tables, \textit{we abbreviated AvgPassRatio as APR}.
More metrics (i.e., Pass@2, Pass@3, and CodeBLEU~\cite{ren2020codebleu}) will be discussed in Section~\ref{sec:threats}.

\subsection{Compared Techniques}
\label{sec:baseline}
To evaluate \tech{} sufficiently, we considered nine typical or state-of-the-art prompting techniques for comparison:
\begin{itemize}[leftmargin=10pt]
    \item \textbf{Zero-shot prompting}~\cite{chen2021evaluating}: directly utilizes the original specification to prompt LLMs for code generation.
    
    \item \textbf{Few-shot prompting}~\cite{chen2021evaluating}: enables LLMs to learn the relationship between specifications and code based on randomly selected $<$specification, code$>$ examples. 
    It concatenates these examples with the original specification to form a prompt, which is then fed to LLMs for code generation.
    
    \item \textbf{CoT prompting}~\cite{wei2022chain}: elicits LLMs to produce intermediate reasoning steps as specification understanding. 
    It incorporates the specification understanding into the original specification to form a prompt, which is then fed to LLMs for code generation.
    Following the existing work~\cite{li2023think,li2023enabling}, we applied CoT prompting in both zero-shot~\cite{kojima2022large} and few-shot~\cite{wei2022chain} settings.
    For ease of presentation, we call them \textbf{Zero-shot-CoT prompting} and \textbf{Few-shot-CoT prompting}.

    \item \textbf{Self-planning prompting}~\cite{jiang2023self}: guides LLMs to decompose the specification into a set of easy-to-solve sub-problems and produce code plan by providing few-shot intent-to-plan examples.
    It incorporates the code plan into the original specification to form a prompt, which is then fed to LLMs for code generation.
    
    \item \textbf{SCoT prompting}~\cite{li2023enabling}: enhances CoT prompting by utilizing program structures (i.e., sequence, branch, and loop structures) to produce intermediate reasoning steps.
    
    \item \textbf{Self-Debugging prompting}~\cite{chen2023teaching}: utilizes the runtime errors and test execution results to guide LLMs for fixing incorrectly generated code.
    
    \item \textbf{Self-Edit prompting}~\cite{zhang2023self}: wraps error messages (produced by test execution) as supplementary comments, including test inputs, expected outputs, actual outputs, and runtime errors.
    Then, the supplementary comments serve as feedback to guide LLMs to fix incorrectly generated code.
    
    \item \textbf{Self-repair prompting}~\cite{olausson2023demystifying}: leverages error messages produced by test execution to enable LLMs to produce a short explanation of why the code failed.
    Then, the explanation guides LLMs to fix the incorrectly generated code.

\end{itemize}

In summary, Zero-shot and Few-shot prompting are typical prompting techniques widely used as baselines in code generation studies~\cite{li2023enabling,li2023acecoder}.
Zero-shot-CoT and Few-shot-CoT are typical thought-eliciting prompting techniques, while Self-planning and SCoT are state-of-the-art techniques in this category.
Self-Debugging, Self-Edit, and Self-repair are state-of-the-art feedback-based prompting techniques.


Note that \tech{} is the first to integrate both thought-eliciting and feedback-based prompting techniques.
For thorough evaluation, we also combined each of the two state-of-the-art thought-eliciting prompting techniques (Self-planning and SCoT) with each of the three state-of-the-art feedback-based prompting techniques (Self-Debugging, Self-Edit, and Self-repair).
For example, SCoT combined with Self-repair (referred to as \textbf{SCoT + Self-repair}) involves applying Self-repair to the code generated by SCoT.
In total, we implemented 15 compared techniques as baselines.
Given the input length limit of LLMs, we used the 4-shot setting for all few-shot baselines following the existing work~\cite{jiang2023self,liu2023verilogeval,jiang2024seed}.
\vspace{-1mm}
\section{Results and Analysis}
\label{sec:results_and_analysis}

\begin{table*}[t!]
    \caption{Effectiveness comparison on ChatGPT in terms of Pass@1 ($\uparrow$) and AvgPassRatio ($\uparrow$). APR is short for AvgPassRatio.}
    \vspace{-2mm}
    \label{tab:rq1-chatgpt}
    \centering
    \tabcolsep=1.4mm
    \begin{adjustbox}{max width=1.0 \textwidth,center}
        \begin{tabular}{ lcccccccccccc }
            \toprule
        	\textbf{ChatGPT} & \multicolumn{2}{c}{\textbf{HumanEval}} & \multicolumn{2}{c}{\textbf{HumanEval${+}$}} & \multicolumn{2}{c}{\textbf{HumanEval-ET}} &
            \multicolumn{2}{c}{\textbf{MBPP-ET}} & \multicolumn{2}{c}{\textbf{APPS}} & \multicolumn{2}{c}{\textbf{APPS-ET}}\\ \cmidrule(lr){2-3} \cmidrule(lr){4-5} \cmidrule(lr){6-7} \cmidrule(lr){8-9} \cmidrule(lr){10-11} \cmidrule(lr){12-13}
        	\textbf{Prompting Technique} & \textbf{Pass@1} & \textbf{APR} & \textbf{Pass@1} & \textbf{APR} & \textbf{Pass@1} & \textbf{APR} & \textbf{Pass@1} & \textbf{APR} & \textbf{Pass@1} & \textbf{APR} & \textbf{Pass@1} & \textbf{APR} \\
        	\midrule
            Zero-shot & 73.78\% & 79.14\% & 66.46\% & 68.66\% & 67.07\% & 82.69\% & 58.21\% & 71.94\% & 16.33\% & 27.98\% & 6.00\% & 23.93\% \\ 
            Few-shot & 75.61\% & 79.77\% & 69.51\% & 70.81\% & 68.29\% & 83.04\% & 59.34\% & 73.34\% & 17.33\% & 24.45\% & 6.33\% & 20.37\% \\ 
            Zero-shot-CoT & 74.39\% & 79.25\% & 67.68\% & 69.69\% & 67.07\% & 83.21\% & 59.14\% & 72.91\% & 18.00\% & 30.02\% & 6.00\% & 24.97\% \\ 
            Few-shot-CoT & 76.83\% & 81.62\% & 70.73\% & 72.03\% & 68.90\% & 84.41\% & 59.34\% & 73.22\% & 20.33\% & 32.65\% & 7.00\% & 27.60\% \\ 
            Self-planning & 78.66\% & 81.95\% & 70.73\% & 72.36\% & 70.73\% & 83.67\% & 62.01\% & 75.22\% & 21.33\% & 33.48\% & 8.33\% & 28.97\% \\ 
            SCoT & 79.27\% & 83.97\% & 71.95\% & 73.66\% & 71.34\% & 85.54\% & 62.22\% & 77.36\% & 22.00\% & 34.07\% & 7.67\% & 29.56\% \\ \midrule
            Self-Debugging & 77.44\% & 81.52\% & 71.34\% & 72.94\% & 70.12\% & 84.18\% & 60.37\% & 75.69\% & 18.67\% & 31.83\% & 6.33\% & 27.36\% \\
            Self-Edit & 76.83\% & 81.15\% & 69.51\% & 70.82\% & 68.29\% & 84.34\% & 61.19\% & 76.35\% & 20.67\% & 33.28\% & 7.00\% & 27.49\% \\
            Self-repair & 80.49\% & 82.61\% & 74.39\% & 75.65\% & 72.56\% & 81.25\% & 64.07\% & 75.25\% & 23.00\% & 30.21\% & 8.00\% & 25.19\% \\  \midrule
            Self-planning+Self-Debugging & 80.49\% & 83.33\% & 72.56\% & 74.20\% & 72.56\% & 85.66\% & 62.01\% & 75.40\% & 21.33\% & 33.67\% & 8.33\% & 29.23\% \\
            Self-planning+Self-Edit & 79.88\% & 83.25\% & 71.34\% & 72.92\% & 71.34\% & 84.10\% & 62.01\% & 75.31\% & 21.33\% & 33.81\% & 8.33\% & 29.36\% \\
            Self-planning+Self-repair & 82.32\% & 84.83\% & 73.78\% & 75.70\% & 73.78\% & 86.56\% & 64.17\% & 75.55\% & 23.33\% & 37.19\% & 8.67\% & 31.65\% \\
            SCoT+Self-Debugging & 81.10\% & 85.50\% & 73.17\% & 74.90\% & 71.95\% & 86.83\% & 63.14\% & 78.45\% & 22.00\% & 34.16\% & 7.67\% & 29.77\% \\
            SCoT+Self-Edit & 79.88\% & 84.27\% & 72.56\% & 74.31\% & 71.95\% & 86.12\% & 63.96\% & 78.58\% & 22.33\% & 34.43\% & 7.67\% & 29.78\% \\
            SCoT+Self-repair & 81.71\% & 84.48\% & 75.00\% & 76.28\% & 73.17\% & 84.66\% & 64.37\% & 79.21\% & 25.00\% & 38.38\% & 8.67\% & 33.05\% \\ 
            \textbf{\tech{}} & \textbf{90.24\%} & \textbf{91.59\%} & \textbf{80.49\%} & \textbf{82.16\%} & \textbf{79.88\%} & \textbf{91.66\%} & \textbf{69.10\%} & \textbf{83.07\%} & \textbf{35.67\%} & \textbf{47.72\%} & \textbf{10.33\%} & \textbf{38.81\%} \\
            \bottomrule
        \end{tabular}
    \end{adjustbox}
    \vspace{-2mm}
\end{table*}

\begin{table*}[t!]
    \caption{Effectiveness comparison on DeepSeek-Coder in terms of Pass@1 ($\uparrow$) and AvgPassRatio ($\uparrow$). APR is short for AvgPassRatio.}
    \vspace{-2mm}
    \label{tab:rq1-deepseek-coder}
    \centering
    \tabcolsep=1.4mm
    \begin{adjustbox}{max width=1.0 \textwidth,center}
        \begin{tabular}{ lcccccccccccc }
            \toprule
        	\textbf{DeepSeek-Coder} & \multicolumn{2}{c}{\textbf{HumanEval}} & \multicolumn{2}{c}{\textbf{HumanEval${+}$}} & \multicolumn{2}{c}{\textbf{HumanEval-ET}} &
            \multicolumn{2}{c}{\textbf{MBPP-ET}} & \multicolumn{2}{c}{\textbf{APPS}} & \multicolumn{2}{c}{\textbf{APPS-ET}}\\ \cmidrule(lr){2-3} \cmidrule(lr){4-5} \cmidrule(lr){6-7} \cmidrule(lr){8-9} \cmidrule(lr){10-11} \cmidrule(lr){12-13}
        	\textbf{Prompting Technique} & \textbf{Pass@1} & \textbf{APR} & \textbf{Pass@1} & \textbf{APR} & \textbf{Pass@1} & \textbf{APR} & \textbf{Pass@1} & \textbf{APR} & \textbf{Pass@1} & \textbf{APR} & \textbf{Pass@1} & \textbf{APR} \\
        	\midrule
            Zero-shot & 76.22\% & 80.92\% & 72.56\% & 73.79\% & 69.51\% & 82.97\% & 55.75\% & 68.75\% & 4.00\% & 11.38\% & 1.00\% & 10.95\% \\ 
            Few-shot & 78.05\% & 82.41\% & 73.17\% & 74.91\% & 70.73\% & 84.28\% & 56.78\% & 69.86\% & 4.33\% & 6.28\% & 1.33\% & 4.81\% \\ 
            Zero-shot-CoT & 76.83\% & 80.86\% & 72.56\% & 73.80\% & 69.51\% & 82.68\% & 55.85\% & 68.78\% & 4.67\% & 11.77\% & 2.00\% & 10.83\% \\ 
            Few-shot-CoT & 78.66\% & 82.48\% & 74.39\% & 75.65\% & 71.34\% & 83.49\% & 57.08\% & 70.55\% & 4.67\% & 10.53\% & 1.67\% & 10.29\% \\ 
            Self-planning & 78.65\% & 82.11\% & 73.17\% & 74.40\% & 70.73\% & 83.45\% & 57.08\% & 70.05\% & 4.00\% & 8.92\% & 1.00\% & 7.64\% \\ 
            SCoT & 78.05\% & 82.03\% & 73.78\% & 75.04\% & 70.73\% & 83.48\% & 58.01\% & 71.16\% & 4.33\% & 10.21\% & 1.33\% & 9.29\% \\ \midrule
            Self-Debugging & 77.44\% & 82.45\% & 73.17\% & 74.78\% & 70.73\% & 83.83\% & 56.88\% & 69.73\% & 4.67\% & 12.11\% & 1.33\% & 11.60\% \\
            Self-Edit & 76.83\% & 81.97\% & 73.17\% & 74.40\% & 70.12\% & 84.32\% & 57.08\% & 71.52\% & 4.67\% & 12.22\% & 1.67\% & 11.70\% \\
            Self-repair & 79.27\% & 83.08\% & 72.56\% & 74.55\% & 70.12\% & 85.34\% & 58.73\% & 73.52\% & 5.67\% & 12.68\% & 1.33\% & 10.41\% \\ \midrule
            Self-planning+Self-Debugging & 79.27\% & 82.72\% & 73.78\% & 75.01\% & 71.34\% & 84.55\% & 59.34\% & 72.50\% & 5.33\% & 11.05\% & 1.67\% & 9.76\% \\
            Self-planning+Self-Edit & 79.88\% & 83.33\% & 74.39\% & 75.62\% & 71.95\% & 84.67\% & 58.83\% & 71.96\% & 5.67\% & 13.14\% & 1.67\% & 12.46\% \\
            Self-planning+Self-repair & 79.88\% & 83.10\% & 74.39\% & 75.62\% & 71.95\% & 84.19\% & 59.34\% & 73.84\% & 6.33\% & 13.91\% & 1.33\% & 11.35\% \\
            SCoT+Self-Debugging & 79.27\% & 83.25\% & 74.39\% & 76.26\% & 71.95\% & 84.70\% & 58.21\% & 71.81\% & 5.67\% & 11.87\% & 1.67\% & 10.88\% \\
            SCoT+Self-Edit & 78.66\% & 82.64\% & 74.39\% & 75.65\% & 70.73\% & 83.91\% & 58.21\% & 71.74\% & 4.67\% & 11.45\% & 1.67\% & 10.56\% \\
            SCoT+Self-repair & 80.49\% & 84.15\% & 73.78\% & 75.77\% & 71.34\% & 86.10\% & 59.45\% & 72.94\% & 6.67\% & 15.47\% & 1.67\% & 15.00\% \\ 
            \textbf{\tech{}} & \textbf{83.54\%} & \textbf{86.17\%} & \textbf{78.66\%} & \textbf{80.22\%} & \textbf{75.00\%} & \textbf{87.12\%} & \textbf{63.35\%} & \textbf{78.03\%} & \textbf{14.00\%} & \textbf{23.59\%} & \textbf{5.00\%} & \textbf{19.81\%} \\
            \bottomrule
        \end{tabular}
    \end{adjustbox}
    \vspace{-1mm}
\end{table*}

\subsection{RQ1: Overall Effectiveness of \tech{}}
\label{sec:rq1}
\subsubsection{Process}
To answer RQ1, we applied \tech{} and 15 compared techniques to ChatGPT and DeepSeek-Coder.
We then measured the effectiveness of each technique on 6 widely-used benchmarks in terms of Pass@1 and AvgPassRatio.

\subsubsection{Results}
Tables~\ref{tab:rq1-chatgpt} and~\ref{tab:rq1-deepseek-coder} show the effectiveness comparison results on ChatGPT and DeepSeek-Coder, respectively. 
We found that SCoT outperforms all thought-eliciting prompting baselines, while Self-repair excels compared to all feedback-based prompting baselines, confirming their effectiveness as state-of-the-art baselines in their corresponding categories.
Also, each combination of thought-eliciting and feedback-based prompting techniques outperforms the corresponding individual techniques, which confirms the synergy between both categories, motivating our \tech{} technique.

In particular, \tech{} achieves the best effectiveness among all studied techniques, demonstrating its stable superiority in both metrics across all subjects (two LLMs with six benchmarks).
Based on ChatGPT, \tech{} significantly improves all compared techniques by 16.02\%$\sim$45.30\% and 11.84\%$\sim$40.03\% in terms of Pass@1 and AvgPassRatio on average across all six benchmarks, respectively.
Based on DeepSeek-Coder, \tech{} significantly improves them by 55.23\%$\sim$114.92\% and 16.83\%$\sim$102.37\% in terms of Pass@1 and AvgPassRatio, respectively.
Furthermore, the \textit{Wilcoxon Signed-Rank Test}~\cite{wilcoxon1970critical} at the significance level of 0.05 confirms that all p-values are smaller than 5.14e-6, demonstrating the statistically significant superiority of \tech{} over all compared techniques.


\begin{table*}[t]
    \caption{Comparison between \tech{} and its variants in terms of Pass@1 ($\uparrow$) and AvgPassRatio ($\uparrow$)}
    \vspace{-2mm}
    \label{tab:rq2}
    \centering
    \tabcolsep=2.3mm
    \begin{adjustbox}{max width=1.0 \textwidth,center}
        \begin{tabular}{ lcccccccccccc }
            \toprule
        	\multirow{2}{*}{\textbf{Variant}} & \multicolumn{2}{c}{\textbf{HumanEval}} & \multicolumn{2}{c}{\textbf{HumanEval${+}$}} & \multicolumn{2}{c}{\textbf{HumanEval-ET}} &
            \multicolumn{2}{c}{\textbf{MBPP-ET}} & \multicolumn{2}{c}{\textbf{APPS}} & \multicolumn{2}{c}{\textbf{APPS-ET}}\\ \cmidrule(lr){2-3} \cmidrule(lr){4-5} \cmidrule(lr){6-7} \cmidrule(lr){8-9} \cmidrule(lr){10-11} \cmidrule(lr){12-13}
        	& \textbf{Pass@1} & \textbf{APR} & \textbf{Pass@1} & \textbf{APR} & \textbf{Pass@1} & \textbf{APR} & \textbf{Pass@1} & \textbf{APR} & \textbf{Pass@1} & \textbf{APR} & \textbf{Pass@1} & \textbf{APR} \\
        	\midrule
            \techVariantWithoutFeedback & 85.37\% & 87.33\% & 76.83\% & 79.11\% & 77.44\% & 88.60\% & 66.32\% & 80.12\% & 27.67\% & 38.63\% & 9.00\% & 32.75\% \\ 
            \techVariantSelfrepair & 85.98\% & 87.94\% & 78.05\% & 79.67\% & 78.05\% & 89.54\% & 68.07\% & 82.38\% & 30.67\% & 42.80\% & 10.00\% & 35.14\% \\
            \techVariantSCoT & 83.54\% & 87.16\% & 76.22\% & 77.87\% & 74.39\% & 87.83\% & 65.81\% & 80.20\% & 26.33\% & 37.72\% & 9.33\% & 32.50\% \\ 
            \techVariantWithoutSelfimprovement & 81.10\% & 84.30\% & 73.78\% & 75.09\% & 72.56\% & 85.31\% & 64.78\% & 79.36\% & 24.33\% & 40.24\% & 8.67\% & 33.58\% \\ 
            \textbf{\tech{}} & \textbf{90.24\%} & \textbf{91.59\%} & \textbf{80.49\%} & \textbf{82.16\%} & \textbf{79.88\%} & \textbf{91.66\%} & \textbf{69.10\%} & \textbf{83.07\%} & \textbf{35.67\%} & \textbf{47.72\%} & \textbf{10.33\%} & \textbf{38.81\%} \\
            \bottomrule
        \end{tabular}
    \end{adjustbox}
    \vspace{-0mm}
\end{table*}

\vspace{-1mm}
\subsection{RQ2: Contribution of Each Main Component in \tech{}} 
\label{sec:rq2}
\subsubsection{Variants}
\tech{} consists of two phases: thought-eliciting and feedback-based prompting phases.
To investigate the contribution of each phase (including some key components in each phase), we created four variants of \tech{}:
\begin{itemize}[leftmargin=10pt]
    \item \textbf{\techVariantWithoutFeedback{}}: we removed the feedback-based prompting phase from \tech{}, i.e., it just uses the specification understanding produced in the thought-eliciting prompting phase to generate code.
    It can measure the effectiveness of the individual thought-eliciting prompting phase in \tech{} and also reflect the contribution of the feedback-based prompting phase.

    \item \textbf{\techVariantSelfrepair{}}: we replaced the feedback-based prompting strategy in \tech{} with the state-of-the-art Self-repair prompting strategy.
    It can investigate the effectiveness of our designed feedback-based prompting strategy.
    
    \item \textbf{\techVariantSCoT{}}: we replaced the thought-eliciting prompting strategy in \tech{} with the state-of-the-art SCoT strategy.
    It can further investigate the effectiveness of our designed thought-eliciting strategy by complementing \techVariantWithoutFeedback{}.
    
    \item \textbf{\techVariantWithoutSelfimprovement{}}: we removed the self-improvement component in the thought-eliciting prompting phase from \tech{}.
    It can investigate the contribution of identifying and \checkedrevision{refining} misunderstanding before code generation in \tech{}.
\end{itemize}

\subsubsection{Process}
Due to the limited computational resource and evaluation time cost, we selected ChatGPT as the representative LLM for the ablation study, as it achieves the best effectiveness with \tech{} (in Tables~\ref{tab:rq1-chatgpt} and~\ref{tab:rq1-deepseek-coder}).
Similarly, we used ChatGPT for the experiments in Section~\ref{sec:discussion}.
Specifically, we applied \tech{} and its four variants to ChatGPT respectively, and measured the effectiveness of each technique on 6 widely-used benchmarks in terms of Pass@1 and AvgPassRatio.

\subsubsection{Results}
Table~\ref{tab:rq2} shows the comparison results among \tech{} and its four variants in terms of Pass@1 and AvgPassRatio.
First, \techVariantWithoutFeedback{} achieves superior effectiveness compared to all 15 baselines in terms of both metrics (except SCoT+Self-repair on APPS-ET in terms of AvgPassRatio).
It demonstrates the effectiveness of our thought-eliciting prompting phase.
\tech{} outperforms \techVariantSCoT{} with average improvements of 12.03\% and 10.74\% in terms of Pass@1 and AvgPassRatio, further confirming the effectiveness of \tech{}'s thought-eliciting prompting strategy, when combining with the same feedback-based prompting strategy (designed in \tech{}).

Second, \tech{} outperforms \techVariantWithoutFeedback{} with average improvements of 10.25\% in Pass@1 and 9.65\% in AvgPassRatio, demonstrating the contribution of our feedback-based prompting phase.
Moreover, \tech{} demonstrates superior effectiveness over \techVariantSelfrepair{} with average improvements of 5.26\% in Pass@1 and 5.40\% in AvgPassRatio, further confirming the superiority of our feedback-based prompting strategy over the state-of-the-art Self-repair strategy, when combining with the same thought-eliciting prompting strategy (designed in \tech{}).

Third, \tech{} demonstrates superior effectiveness compared to \techVariantWithoutSelfimprovement{} with average improvements of 17.15\% and 10.72\% in terms of Pass@1 and AvgPassRatio.
In addition to the fixing process in the feedback-based phase, \tech{} also performs another misunderstanding fixing through the self-improvement process in the thought-eliciting phase.
The results confirm the necessity of the self-improvement process in \tech{}.

Furthermore, the \textit{Wilcoxon Signed-Rank Test}~\cite{wilcoxon1970critical} at the significance level of 0.05 confirms that all p-values are smaller than 1.39e-4, demonstrating the statistically significant superiority of \tech{} over all variants.
Overall, each main component does make contributions to the overall effectiveness of \tech{}.

\vspace{-1mm}
\begin{table*}[t]
    \caption{\checkedrevision{Influence of Hyper-parameters N and M in terms of Pass@1 ($\uparrow$) and AvgPassRatio ($\uparrow$)}}
    \vspace{-2mm}
    \label{tab:n_m}
    \centering
    \tabcolsep=1.25mm
    \begin{adjustbox}{max width=1.0 \textwidth,center}
        \begin{tabular}{ llcccccccccccc }
            \toprule
        	\multirow{2}{*}{\revision{\textbf{LLM}}} & \multirow{2}{*}{\revision{\textbf{Configuration}}} & \multicolumn{2}{c}{\revision{\textbf{HumanEval}}} & \multicolumn{2}{c}{\revision{\textbf{HumanEval${+}$}}} & \multicolumn{2}{c}{\revision{\textbf{HumanEval-ET}}} &
            \multicolumn{2}{c}{\revision{\textbf{MBPP-ET}}} & \multicolumn{2}{c}{\revision{\textbf{APPS}}} & \multicolumn{2}{c}{\revision{\textbf{APPS-ET}}}\\ \cmidrule(lr){3-4} \cmidrule(lr){5-6} \cmidrule(lr){7-8} \cmidrule(lr){9-10} \cmidrule(lr){11-12} \cmidrule(lr){13-14}
        	& & \revision{\textbf{Pass@1}} & \revision{\textbf{APR}} & \revision{\textbf{Pass@1}} & \revision{\textbf{APR}} & \revision{\textbf{Pass@1}} & \revision{\textbf{APR}} & \revision{\textbf{Pass@1}} & \revision{\textbf{APR}} & \revision{\textbf{Pass@1}} & \revision{\textbf{APR}} & \revision{\textbf{Pass@1}} & \revision{\textbf{APR}} \\
        	\midrule
            \multirow{2}{*}{\revision{ChatGPT}} & \revision{\tech{}$_{N=1,M=1}$} & \revision{90.24\%} & \revision{91.59\%} & \revision{80.49\%} & \revision{82.16\%} & \revision{79.88\%} & \revision{91.66\%} & \revision{69.10\%} & \revision{83.07\%} & \revision{35.67\%} & \revision{47.72\%} & \revision{10.33\%} & \revision{38.81\%} \\
            & \revision{\tech{}$_{N=2,M=2}$} & \revision{90.85\%} & \revision{92.88\%} & \revision{81.10\%} & \revision{82.74\%} & \revision{80.49\%} & \revision{92.04\%} & \revision{69.61\%} & \revision{83.32\%} & \revision{39.00\%} & \revision{49.81\%} & \revision{11.33\%} & \revision{40.43\%} \\ \midrule
            \multirow{2}{*}{\revision{DeepSeek-Coder}} & \revision{\tech{}$_{N=1,M=1}$} & \revision{83.54\%} & \revision{86.17\%} & \revision{78.66\%} & \revision{80.22\%} & \revision{75.00\%} & \revision{87.12\%} & \revision{63.35\%} & \revision{78.03\%} & \revision{14.00\%} & \revision{23.59\%} & \revision{5.00\%} & \revision{19.81\%} \\
            & \revision{\tech{}$_{N=2,M=2}$} & \revision{84.76\%} & \revision{86.98\%} & \revision{79.88\%} & \revision{81.44\%} & \revision{76.22\%} & \revision{88.35\%} & \revision{65.71\%} & \revision{79.59\%} & \revision{16.33\%} & \revision{27.31\%} & \revision{5.33\%} & \revision{22.84\%} \\
            \bottomrule
        \end{tabular}
    \end{adjustbox}
    \vspace{-0mm}
\end{table*}

\subsection{\checkedrevision{RQ3: Influence of Hyper-parameters}} 
\label{sec:rq3}
\subsubsection{\checkedrevision{Setup}}
\checkedrevision{
\tech{} involves three main hyper-parameters: $N$ (the number of refinements), $M$ (the number of adjustments), and the decoding temperature of LLMs (denoted as $T$). 
By default, we set $N$=$1$ and $M$=$1$ to balance the effectiveness and efficiency.
Following the existing work~\cite{wang2022self,wan2023universal}, we set $T$=$0.7$ for both LLMs.
In this RQ, we investigated the performance of \tech{} under different settings.
Due to the evaluation cost, we considered $N$=$\{1, 2\}$, $M$=$\{1, 2\}$, and $T$=$\{0.6, 0.7, 0.8, 0.9, 1.0\}$.
}

\subsubsection{\checkedrevision{Results}}
\checkedrevision{
Table~\ref{tab:n_m} shows that increasing $N$ and $M$ enhances the effectiveness of \tech{}.
Specifically, \tech{}$_{N=2,M=2}$ outperforms \tech{}$_{N=1,M=1}$ with the average improvements of 4.46\% and 4.03\% in terms of Pass@1 and AvgPassRatio, respectively, across all the six benchmarks and both LLMs.
However, larger $N$ and $M$ also incur more time and token costs.
\tech{}$_{N=2,M=2}$ takes 44.54\% more time and 40.94\% more tokens than \tech{}$_{N=1,M=1}$.
We will discuss the efficiency of our studied techniques in detail in Section~\ref{sec:codet}.
Hence, we used $N$=$1$ and $M$=$1$ as the default setting, as \tech{} under this setting consistently outperforms all baselines across all benchmarks and both LLMs and has lower time and token costs than other settings with larger $N$ and $M$, demonstrating the cost-effectiveness of \tech{} under this setting.
}

\checkedrevision{
We then investigated the influence of decoding temperature.
Due to the space limit, we put the detailed results on our homepage~\cite{muFiX2023}.
For example, \tech{} under all the studied settings for the decoding temperature, consistently outperforms the state-of-the-art SCoT + Self-repair, by achieving 16.02\%$\sim$22.60\% higher Pass@1 and 11.84\%$\sim$24.36\% higher AvgPassRatio averaging across all six benchmarks on ChatGPT.
This demonstrates the stable superiority of \tech{} under different decoding-temperature settings.
}
\vspace{-1mm}
\section{Discussion}
\label{sec:discussion}
\subsection{Efficiency}
\label{sec:codet}
\checkedrevision{We measured the time and token overhead on code generation. 
On average, the most efficient baseline (Zero-shot prompting) takes 5.78s for a programming task, while the most effective baseline (SCoT + Self-repair) and \tech{} take 20.61s and 22.75s, respectively.
The average token overheads for Zero-shot, SCoT + Self-repair, and \tech{} are 0.29K, 3.31K, and 3.44K, respectively.
The time and token overhead of \tech{} are slightly higher than those of SCoT + Self-repair.}
This is because \tech{} involves a bit more LLM invocations compared to existing prompting techniques.
For example, the state-of-the-art thought-eliciting technique (SCoT) involves two LLM invocations for a programming task, whereas our thought-eliciting strategy involves four invocations.
Considering the significant effectiveness of \tech{}, some extra cost is acceptable, illustrating its excellent balance of cost and effectiveness.

Recent studies~\cite{chen2022codet,tian2022learning,li2023enabling} suggest that increasing LLM invocations can enhance code generation performance.
Hence, we further conducted a comparison experiment under the same number of LLM invocations to demonstrate \tech{}'s effectiveness more clearly.
CodeT~\cite{chen2022codet} is the state-of-the-art method to improve code generation by increasing LLM invocations. 
It generates more candidate code instances and ranks them by test execution.
Such ranking strategies can be used to enhance thought-eliciting prompting due to their orthogonal effect~\cite{li2023acecoder,li2023enabling}.
Hence, we combined CodeT with the state-of-the-art thought-eliciting technique (SCoT) to compare with \techVariantWithoutFeedback{} under the same number of LLM invocations.
Both techniques do not involve feedback-based prompting for a fair comparison.
We used ChatGPT as the representative LLM.

Due to the space limit, we put detailed results on our homepage~\cite{muFiX2023} and summarized the conclusions here.
\techVariantWithoutFeedback{} achieves 11.45\% higher Pass@1 and 6.93\% higher AvgPassRatio than SCoT averaging across all six benchmarks.
SCoT + CodeT can also outperform SCoT, but just achieve 3.65\% higher Pass@1 and 3.55\% higher AvgPassRatio.
The results clearly demonstrate the effectiveness of our \tech{}.

\begin{table}[t]
    \vspace{-2mm}
    \caption{Influence of test cases used in \tech{}}
    \vspace{-2mm}
    \label{tab:testcase_number}
    \centering
    \tabcolsep=0.9mm
    \begin{adjustbox}{max width=1.0 \textwidth,center}
        \begin{tabular}{ ccccccc }
            \toprule
        	\multirow{2}{*}{\textbf{Number}} & \multicolumn{2}{c}{\textbf{HumanEval}} & \multicolumn{2}{c}{\textbf{HumanEval${+}$}} & \multicolumn{2}{c}{\textbf{HumanEval-ET}} \\ \cmidrule(lr){2-3} \cmidrule(lr){4-5} \cmidrule(lr){6-7} 
        	& \textbf{Pass@1} & \textbf{APR} & \textbf{Pass@1} & \textbf{APR} & \textbf{Pass@1} & \textbf{APR} \\
        	\midrule
            SCoT+Self-repair & 81.71\% & 84.48\% & 75.00\% & 76.28\% & 73.17\% & 84.66\% \\ \hdashline
            \techVariantWithoutTestcase & 86.00\% & 88.66\% & 78.05\% & 80.17\% & 78.66\% & 90.47\%  \\
            \hdashline
            1 & 82.32\% & 85.98\% & 75.61\% & 77.19\% & 73.17\% & 87.32\% \\
            2 & 85.37\% & 87.04\% & 77.44\% & 79.12\% & 76.22\% & 87.52\% \\
            3+ & \textbf{90.24\%} & \textbf{91.59\%} & \textbf{80.49\%} & \textbf{82.16\%} & \textbf{79.88\%} & \textbf{91.66\%} \\
            \bottomrule
        \end{tabular}
    \end{adjustbox}
    \vspace{-0mm}
\end{table}

\subsection{Influence of Test Cases}
\label{sec:influence_testcase}
First, we conducted an experiment to investigate the influence of the number of test cases used in \tech{}. 
We set the number of test cases to \{1, 2, 3+\}, with the notation 3+ indicating the utilization of all available test cases in a programming specification (the default setting in \tech{}).
Note that we randomly selected the corresponding numbers of test cases from the whole set of test cases in each programming specification.
If fewer test cases in the specification were available than desired, we used all available ones.
We used ChatGPT as the representative LLM on three benchmarks (HumanEval, HumanEval$+$, and HumanEval-ET).
Table~\ref{tab:testcase_number} shows the effectiveness of \tech{} with different numbers of test cases in terms of Pass@1 and AvgPassRatio.
\tech{} always exhibits better performance than the most effective baseline (SCoT + Self-repair), even with only one test case. 
As the number of test cases increased, \tech{}'s code generation performance improved, demonstrating the importance of test-case-driven specification misunderstanding fixing for code generation. 

\checkedrevision{Second, although it is quite common that a few test cases are included as part of a specification, \ul{we need to consider the scenario when test cases are absent, further enhancing the generality of \tech{}}.
To prove the concept, we utilized LLMs to automatically generate test cases for each programming specification (without test cases) following the existing work~\cite{chen2022codet,liu2023your}.
We call this variant \techVariantWithoutTestcase{}.
We investigated the effectiveness of \techVariantWithoutTestcase{} by taking ChatGPT as the representative model on the same three benchmarks as above.
We set the number of generated test cases to 3 as most of the programming specifications include three test cases as mentioned in Section~\ref{sec:benchmarks}.
From Table~\ref{tab:testcase_number}, \techVariantWithoutTestcase{} significantly outperforms the most effective baseline (SCoT + Self-repair) by 5.61\% and 5.64\% in terms of Pass@1 and AvgPassRatio, respectively,
which demonstrates the practical effectiveness of \tech{} in such a scenario.
Also, test-case quality could affect the effectiveness of \tech{}. 
We manually analyzed some cases where \techVariantWithoutTestcase{} underperformed \tech{} and found that LLM-generated test cases covered only one branch mentioned in the specification or even contained errors, leading to slightly worse effectiveness. 
In the future, we will design better test generation methods, such as incorporating test coverage to guide LLM-based test generation, to further enhance the effectiveness of \tech{}.}
\checkedrevision{Besides, the resources may be limited in practice. 
In such scenarios, employing test prioritization to select higher-quality test cases may help ensure the effectiveness of \tech{}.
We regard this promising exploration as our future work.}

\subsection{\checkedrevision{Comparison with Agent-based Techniques}}
\label{sec:agent_based_technique}
\checkedrevision{
Recently, some agent-based techniques leveraging interactions among agents have been proposed to enhance LLMs' effectiveness in software development.
They are actually orthogonal to \tech{} to a large extent, and we could utilize \tech{} to enhance individual agents for better code generation.
Nevertheless, we still investigated 4 state-of-the-art agent-based techniques (MetaGPT~\cite{hong2023metagpt}, Self-collaboration~\cite{dong2023self}, ChatDev~\cite{qian2023communicative}, and AgentCoder~\cite{huang2023agentcoder}) for comparison.
They share the same high-level insight, which simulates the software development process by assigning roles to several LLM agents for collaboration in code generation.
The difference among them mainly lies in that they assign different roles to LLM agents and design different strategies to solve the corresponding tasks.
Due to the limited space, more details about these techniques can be found in their corresponding papers.
}

    
    
    
    

\checkedrevision{
We used ChatGPT as the representative in this experiment due to the evaluation cost.
Note that we used the default numbers of iterations for MetaGPT, Self-collaboration, ChatDev, and AgentCoder in this experiment, which are 3, 4, 10, and 5, respectively, and are all larger than the single iteration of \tech{}.
While this comparison may be unfavorable to \tech{}, \tech{} still outperforms the four agent-based techniques by 19.23\%$\sim$37.20\% and 13.48\%$\sim$60.33\% in terms of Pass@1 and AvgPassRatio, respectively, averaging across all the six benchmarks.
Due to space limit, we put the detailed results on our project homepage~\cite{muFiX2023}.
The \textit{Wilcoxon Signed-Rank Test}~\cite{wilcoxon1970critical} at the significance level of 0.05 confirms that all p-values are smaller than 1.38e-3, demonstrating the statistically significant superiority of \tech{} over all the studied agent-based techniques.
Particularly, as demonstrated in Section~\ref{sec:rq3}, \tech{}'s effectiveness can be further enhanced with additional iterations (i.e., increasing $N$ and $M$).
Among the four agent-based techniques, AgentCoder is the most effective and efficient but still takes an average of 211.75 seconds and 63.77K tokens per programming task, consuming 830.77\% more time and 1753.78\% more tokens than \tech{}.
Overall, \tech{} exhibits significant superiority over these agent-based techniques.
}

\subsection{\checkedrevision{Soundness of Our Checking Mechanism}}
\label{sec:checking_method}
\checkedrevision{Achieving optimality in semantic analysis is generally undecidable. 
Many existing techniques, e.g., search-based solutions in software engineering, are known to suffer from local optimum~\cite{kulal2019spoc}. 
The uninterpretable nature of LLMs exacerbates this difficulty. 
Hence, there does not exist a decision procedure to check if an LLM correctly understands a specification, rendering sound checking mechanisms for LLMs' understanding intractable.
Such similar limitations are general for all deep-learning-based methods.
To relieve this challenge, in \tech{}, we employed a checking mechanism \textit{on the downstream code generation task}. 
Specifically, we checked the correctness of the generated code during the feedback-based prompting phase through the actual execution of test cases inherently included in specifications.
That is, the improvement of understanding is demonstrated by the better effectiveness in downstream code generation.
Our study indeed confirms that this mechanism helps improve the effectiveness of code generation, empirically demonstrating that leveraging existing test cases in specifications can alleviate LLM hallucinations.
}

\subsection{\checkedrevision{Exploration of LLM Combinations in \tech{}}}
\label{sec:different_llms_refinement}
\checkedrevision{We explored the effectiveness of using different LLMs for understanding refinement and code generation via a preliminary experiment.
Due to the inherent characteristics, DeepSeek-Coder's reasoning capability is weaker than ChatGPT's. 
When we used ChatGPT for understanding refinement and DeepSeek-Coder for code generation, the improvements are 1.55\% and 1.01\% in terms of Pass@1 and AvgPassRatio, respectively, averaging across HumanEval, HumanEval${+}$, and HumanEval-ET, compared to using DeepSeek-Coder for both.
This indicates the potential of \tech{} with more appropriate LLMs in corresponding aspects, which can be regarded as our future work.
}

\section{Threats to Validity}
\label{sec:threats}
\vspace{-1mm}
The first threat lies in the generalizability of experimental results. 
To mitigate this threat, we comprehensively selected benchmarks, metrics, baselines, and LLMs. 
Following previous studies~\cite{jiang2023self,dong2023codescore,li2023enabling,li2023think}, we selected six widely-used benchmarks in code generation and employed two metrics for code correctness assessment.
Besides, we used Pass@2, Pass@3, and CodeBLEU~\cite{ren2020codebleu} for measuring code generation performance (although CodeBLEU suffers from some limitations~\cite{eghbali2022crystalbleu,dong2023codescore,li2023enabling}), and put the results on our homepage~\cite{muFiX2023} due to space limitation.
The conclusions are consistent.
We also selected 9 typical or state-of-the-art prompting techniques (as well as 6 combined techniques) for comparison and conducted a comprehensive evaluation on two state-of-the-art LLMs (i.e., ChatGPT and DeepSeek-Coder).
In the future, we will evaluate \tech{} to improve LLMs' code generation performance on more comprehensive benchmarks.

The second threat lies in the randomness involved in LLMs.
On one hand, our large-scale study and consistent conclusions across all subjects can help reduce this threat.
\checkedrevision{On the other hand, we repeated the experiment comparing \tech{} with the most effective baseline (SCoT + Self-repair) on three benchmarks (HumanEval, HumanEval${+}$, and HumanEval-ET) as the representative for three times due to the huge evaluation cost.
The standard derivations for SCoT + Self-repair and \tech{} are only 0.021 and 0.009 for Pass@1, demonstrating the robustness of our conclusions to a large extent.
This further reduces the threat.
Due to the limited space, we place the detailed results on our project homepage~\cite{muFiX2023}.
}

The third threat lies in the design of prompts in \tech{}.
We did not specially tune the natural-language-description format of prompts, and thus cannot ensure their optimality. 
We designed some similar prompts via paraphrasing (e.g., synonym replacement and active-passive sentence transformation), which did not affect \tech{}'s effectiveness much. 
We will systematically investigate \tech{}'s robustness in the future. 
Furthermore, the prompting structure in \tech{} mainly emphasizes the inputs and outputs of test cases, which are essential test-case elements regardless of specifications, also demonstrating generalizability.
\section{Related Work}
\label{sec:related}
Prompting techniques have been demonstrated effective to improve the code generation performance of LLMs in a plug-and-play manner~\cite{wei2022chain,dong2023self,liu2023improving,nashid2023retrieval,jiang2023self}.
In general, they can be divided into two main categories: thought-eliciting prompting techniques and feedback-based prompting techniques.
The former aims to elicit LLMs to produce intermediate reasoning steps for more accurate code generation.
We have introduced and compared CoT~\cite{wei2022chain}, Self-planning~\cite{jiang2023self}, and SCoT~\cite{li2023enabling} techniques in Section~\ref{sec:evaluation_design}.
Besides, KPC~\cite{ren2023misuse} is a knowledge-driven prompting technique, which decomposes code generation into intermediate reasoning steps and utilizes fine-grained knowledge extracted from API documentation for code generation (especially in exception-handling code).
The latter category of prompting techniques use error messages produced by test execution to enable LLMs to fix incorrectly generated code,
such as SED~\cite{gupta2020synthesize}, CodeRL~\cite{le2022coderl}, Self-Debugging~\cite{chen2023teaching}, Self-Edit~\cite{zhang2023self}, and Self-repair~\cite{olausson2023demystifying}.
Due to the evaluation cost, we did not select all these prompting techniques in our study but just selected some typical or state-of-the-art ones as baselines.

Different from the existing techniques, \tech{} is the first to explore the synergy of both categories by devising both sophisticated thought-eliciting and feedback-based prompting.
Its core of improving LLMs' code generation performance is to fix LLMs' specification misunderstanding in each phase.


\section{Conclusion}
In this work, we propose a novel prompting technique \tech{} to improve the code generation performance of LLMs.
Different from the existing prompting techniques, \tech{} devises both sophisticated thought-eliciting prompting and feedback-based prompting, and explores their synergy.
Our thought-eliciting prompting strategy in \tech{} exploits test case analysis and the misunderstanding fixing process to obtain \checkedrevision{more accurate specification understanding}.
Then, our feedback-based prompting strategy further fixes the understanding to reduce the gap between \checkedrevision{the provided refined understanding} (from the first phase) and the actual understanding implicitly utilized by LLMs for code generation.
We conducted an extensive study on two advanced LLMs with six widely-used benchmarks, demonstrating the superiority of \tech{} over the state-of-the-art prompting techniques.

\section*{Acknowledgements}
This work was supported by the National Key Research and Development Program of China (Grant No. 2024YFB4506300), and the National Natural Science Foundation of China (Grant Nos. 62322208, 12411530122).


\balance
\bibliographystyle{IEEEtran}
\bibliography{reference}

\end{document}